\documentclass[aps, prd, showkeys, twocolumn, superscriptaddress, nofootinbib, floatfix]{revtex4-2}

\usepackage{amsmath,amssymb}
\usepackage{graphicx}
\usepackage{dcolumn}
\usepackage{bm}

\usepackage[bookmarksnumbered, pdfpagelabels=true, plainpages=false, colorlinks=true, linkcolor=blue, citecolor=blue, urlcolor=blue]{hyperref}

\usepackage{url}
\usepackage{xcolor}
\usepackage[T1]{fontenc}
\usepackage{amssymb}
\usepackage{subcaption}

\begin{document}

\title{Radial Oscillations of Viscous Stars at Finite Temperature}

\author{Amanda Guerrieri}
\email{amguerrieri@cbpf.br}
\affiliation{CBPF -- Centro Brasileiro de Pesquisas F\'{\i}sicas, 22290-180, Rio de Janeiro, RJ, Brazil.}

\author{Gabriel S. Rocha}
\email{gabrielsr@id.uff.br}
\affiliation{Instituto de F\'{\i}sica, Universidade Federal Fluminense, Niter\'{o}i, Rio de Janeiro, 24210-346, Brazil}

\author{Gabriel S. Denicol}
\email{gsdenicol@id.uff.br}
\affiliation{Instituto de F\'{\i}sica, Universidade Federal Fluminense, Niter\'{o}i, Rio de Janeiro, 24210-346, Brazil}

\author{Raissa F. P. Mendes}
\email{rfpmendes@id.uff.br}
\affiliation{Instituto de F\'{\i}sica, Universidade Federal Fluminense, Niter\'{o}i, Rio de Janeiro, 24210-346, Brazil}
\affiliation{CBPF -- Centro Brasileiro de Pesquisas F\'{\i}sicas, 22290-180, Rio de Janeiro, RJ, Brazil.}

\begin{abstract}
We study the radial oscillation spectrum of relativistic stars within Israel–Stewart and Navier–Stokes theories, extending previous analyses to include heat diffusion and a thermodynamically consistent finite-temperature equation of state. The inclusion of heat flux gives rise to a distinct thermal sector in the mode spectrum, whose structure closely mirrors the dispersion relations of an infinite dissipative fluid. Within Israel–Stewart theory, the thermal modes transition from purely damped to propagating behavior above a critical overtone number, providing a finite-size realization of relativistic second sound in compact stars. Remarkably, the finite stellar geometry can push even the fundamental thermal mode into the propagating regime — a feature with no continuum analogue. For the class of equations of state considered here, where finite-temperature corrections enter as controlled, Sommerfeld-type perturbations of a cold polytrope, the thermal sector couples only weakly to the ordinary fluid oscillation spectrum, with the coupling being of second order in a suitable temperature parameter. We further show that the discrete stellar spectrum is well captured by an analytic ansatz constructed from the flat-spacetime dispersion relations, with the star's finite radius discretizing the continuous mode structure. Our results complete the analysis of radial oscillations of viscous stars by incorporating the last remaining dissipative degree of freedom within the Israel–Stewart framework.
\end{abstract}

\maketitle

\section{Introduction}

Neutron star oscillations are promising sources of gravitational waves for next-generation observatories, such as Cosmic Explorer \cite{Reitze:2019iox} and the Einstein Telescope \cite{Abac_2026}. Once detected, they may offer a unique probe of fundamental physics across multiple fronts \cite{Kokkotas:1999bd,Nollert:1999ji,Pratten:2019sed}. For instance, the oscillation spectrum can encode signatures of new physics, including possible deviations from general relativity \cite{Silva:2024cit,Doneva:2022ewd}. Most notably, however, it is sensitive to the microphysics of neutron stars, providing valuable constraints on the equation of state (EoS) of nuclear matter at supranuclear densities \cite{Dietrich:2020eud,Shibata:2005xz,Hotokezaka:2011dh,Bauswein:2011tp,Bauswein:2012ya,Bauswein:2014qla}. In addition, neutron star oscillations may carry imprints of transport properties \cite{Alford:2017rxf,Alford:2018lhf,Most:2021zvc}, thereby opening a window into the dissipative dynamics of dense nuclear matter.

Recently, there has been growing effort to incorporate viscous effects both in numerical relativity simulations of neutron star nonlinear dynamics \cite{Pandya:2022pif, Pandya:2022sff,Bantilan:2022ech, Camelio:2022fds, Camelio:2022ljs, Chabanov:2023abq,Chabanov:2023blf,Shum:2025jnl,Clarisse:2025lli} and in analyses of their linear oscillations \cite{Redondo-Yuste:2024vdb,Caballero:2025omv,Keeble:2026bzo,Bussieres:2026rnz}. In particular, the authors have contributed to this program by studying the radial oscillation spectrum of viscous stars in Ref.~\cite{Mendes:2025oib} (hereafter, Paper I), under the assumption of a vanishing diffusion current—i.e., including bulk and shear viscosity while neglecting heat conduction. 

The analysis in Paper I was carried out within two models of dissipative hydrodynamics: relativistic Navier–Stokes theory and a version of Israel–Stewart (IS) theory \cite{Israel:1976tn,Israel:1979wp}. The former posits constitutive relations for dissipative currents that are first order in gradients of the hydrodynamic variables; however, it is acausal and known to suffer from instabilities \cite{hiscock:85generic,pichon:65etude}. These issues are remedied in IS theory through the introduction of relaxation-type equations for the dissipative currents \cite{hiscock1983stability,Olson1990EnergyFrame,Bemfica:2017wps}, which makes it of second-order in the hydrodynamic variables. Despite its shortcomings, it is worth mentioning that Navier–Stokes theory provides the unique set of constitutive relations at first order in a gradient expansion, meaning that it can still provide a useful leading-order approximation in regimes where gradients are sufficiently small and relaxation effects are subdominant. 

\begin{figure}[t] 
    \includegraphics[width=\linewidth]{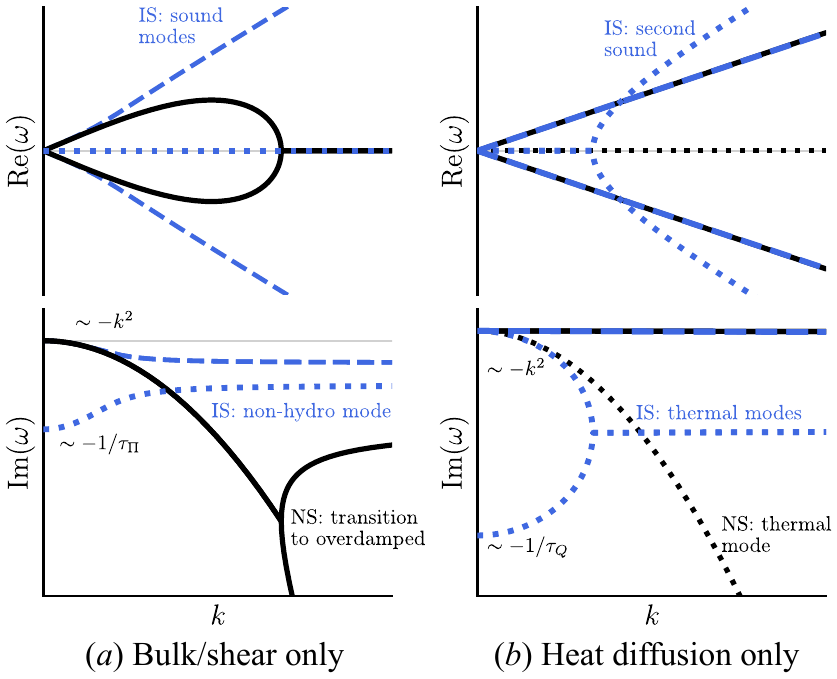}
    \caption{\textit{Schematic dispersion relations for dissipative relativistic fluids}, with details depending on the choice of transport coefficients. The top (bottom) panels display the real (imaginary) part of the mode frequencies as functions of the wavenumber $k$. Panel (a) includes only bulk (or shear) viscosity. There, black curves correspond to Navier–Stokes (NS) theory, where sound modes become overdamped beyond a critical $k$, while blue curves correspond to Israel–Stewart (IS) theory, which additionally contains a purely damped non-hydrodynamic branch (dotted) associated with viscous relaxation. Panel (b) includes only heat diffusion, which adds a thermal branch in the NS case and an additional non-hydrodynamic branch in the IS case (thermal branches are represented as dotted). In IS theory, thermal modes may transition from purely damped to propagating beyond a critical $k$. The spectrum of viscous stars shares the main qualitative features of this continuum picture, but becomes discrete due to the star’s finite size.}
    \label{fig:intro_modes}
\end{figure}

In Paper I, it was shown that the mode structure of viscous stars is broadly consistent with the analysis of longitudinal perturbations around a homogeneous equilibrium state in Minkowski spacetime, depicted pictorially in Fig.~\ref{fig:intro_modes}(a). In the continuum description, Navier–Stokes theory predicts, for each wavenumber $k$, a pair of modes that are propagating (sound-like) at low $k$ but become overdamped (non-propagating) beyond a critical wavenumber. In contrast, Israel–Stewart theory admits additional non-hydrodynamic modes associated with bulk and shear viscous degrees of freedom. At low viscosities, these modes are purely damped, with decay times set by the corresponding relaxation timescales. Paper I showed that these qualitative features persist in viscous stars, the main difference being that the stellar spectrum is discrete ($k \to k_n, n\in\mathbb{N}$) due to the star’s finite size. At the same time, the inhomogeneous background and nontrivial gravitational field introduce genuinely new effects, most notably the crucial role of the fundamental bulk-viscous non-hydrodynamic mode in determining stability against gravitational collapse \cite{Mendes:2025oib}.


An important simplifying assumption in Paper I was the neglect of heat flux. From a technical standpoint, this simplifies the problem, allowing it to be formulated in the frequency domain as a single master equation for the Lagrangian displacement. From a physical perspective, it avoids the need to introduce finite-temperature corrections to the nuclear EoS, which is often modeled as cold (i.e., barotropic) in linear perturbation analyses. The present work aims to complete the analysis of radial oscillations of viscous stars by considering a finite-temperature EoS and including perturbations associated with heat diffusion.

The inclusion of heat diffusion is not only of intrinsic interest, but also relevant in light of recent studies of relativistic stellar dynamics based on alternative formulations of dissipative hydrodynamics, such as the BDNK framework \cite{Bemfica:2017wps, Bemfica:2019knx, Bemfica:2020zjp, Kovtun:2019hdm}. In these approaches, heat flow appears generically as a consequence of adopting hydrodynamic frames different from the standard Eckart or Landau prescriptions \cite{Eckart:1940te,landau:59fluid}. Motivated in part by these developments, in the present work we investigate the role of heat diffusion within the more traditional Navier–Stokes and Israel–Stewart theories.

Our results show that the inclusion of heat diffusion gives rise to a distinct thermal sector in the oscillation spectrum of viscous stars. Within Israel–Stewart theory, this sector contains both hydrodynamic and non-hydrodynamic thermal modes, whose behavior closely parallels the dispersion relations of perturbations around homogeneous equilibrium -- see Fig.~\ref{fig:intro_modes}(b). In particular, we find that thermal modes may transition from purely damped to propagating behavior above a critical overtone number, providing a finite-size realization of relativistic second sound\footnote{Throughout this work we use the terminology ``second sound'' for thermal disturbances that propagate as damped waves instead of showing a diffusive spreading \cite{Joseph:1989zz,Andersson:2011fd}. It differs from ordinary (``first'') sound in that the latter is a propagating disturbance of pressure and density. In Israel-Stewart theory the second-sound behavior originates from the finite relaxation time of the heat flux, which renders the heat equation hyperbolic.} in compact stars, i.e. propagating thermal waves supported by the relaxation dynamics of the heat flux .
Moreover, the finite stellar geometry imposes a minimum wavelength on perturbations, allowing even the fundamental thermal mode to become propagating for certain choices of transport coefficients — a feature absent in the continuum limit, where arbitrarily long wavelengths always access the purely diffusive regime. Despite the richness of this thermal sector, we find that it couples only weakly to the ordinary fluid oscillation spectrum, with the coupling suppressed as $O (\chi^2)$, where $\chi = T/\mu \ll 1$ is the chemical-potential rescaled temperature. The remainder of this work details the analysis underlying these results.

The paper is organized as follows. In Sec.~\ref{sec:framework}, we present the theoretical framework adopted in this work, including the dissipative hydrodynamic models, equation of state, and transport coefficients. In Sec.~\ref{sec:perturbations}, we first review relevant results for linear perturbations of an infinite homogeneous fluid in Minkowski spacetime, which later help interpret our numerical results, and then derive the perturbation equations governing radial oscillations of static, spherically symmetric backgrounds. Numerical results for the heat-flux sector, as well as for the coupled heat-flux and bulk-viscous sectors, are presented in Sec.~\ref{sec:results}. Finally, Sec.~\ref{sec:conclusions} summarizes our main conclusions. Unless stated otherwise, we use units such that $c = G = k_B = 1$.


\section{Framework}
\label{sec:framework}


\subsection{Models of dissipative hydrodynamics and thermodynamic consistency}
\label{sec:hydro-models}

Relativistic hydrodynamics is formulated in terms of local conservation laws for
energy--momentum and conserved charges, supplemented by constitutive
relations constrained by relativistic covariance and thermodynamic
consistency. The continuity equations related to energy-momentum and
net-charge conservation are%
\begin{subequations} \label{eq:CL}
\begin{align}
\label{eq:CL-emt}
\nabla _{\mu }T^{\mu \nu } &= 0, \\
\label{eq:CL-d4c}
\nabla _{\mu }N^{\mu } &= 0,
\end{align}
\end{subequations}
with $N^{\mu }$ being the net-baryon 4-current and $T^{\mu \nu }$ being the
energy-momentum tensor. 

To make contact with fluid dynamics, these conserved currents are decomposed with respect to a time-like, normalized 4-velocity $u^\mu$ \cite{degroot1980,Denicol:2021,Rezzolla:2013dea},
\begin{subequations}
\begin{align}
T^{\mu \nu } &= \mathcal{E} u^{\mu }u^{\nu }+ \mathcal{P} \Delta ^{\mu \nu
}+Q^{\mu }u^{\nu }+Q^{\nu }u^{\mu }+\pi ^{\mu \nu }, \\
N^\mu &= \mathcal{N} u^{\mu } +\mathcal{J}^{\mu } ,
\end{align}
\end{subequations}
where we defined the projection operator $\Delta^{\mu \nu} \equiv g^{\mu\nu} + u^\mu u^\nu$. This decomposition introduces the fundamental hydrodynamic variables: the local
energy density $ \mathcal{E}$, the isotropic pressure $\mathcal{P} $, the heat flow 4-current $Q^{\mu }$, the shear stress tensor $\pi^{\mu \nu }$, the local net-charge density $\mathcal{N}$ and the net-charge diffusion 4-current $\mathcal{J}^{\mu }$.

The conservation laws alone do not form a closed system and must be supplemented by constitutive relations determining the form of the currents, together with an equation of state that specifies the thermodynamic properties of the system. When gravitational effects are relevant, these equations must be solved consistently together with Einstein’s equations, 
\begin{equation}
\label{eq:EFE}
G_{\mu \nu} \equiv R_{\mu \nu} - \frac{1}{2} g_{\mu \nu} R  = 8 \pi T_{\mu \nu},    
\end{equation}
which determine the spacetime geometry.

\subsubsection{Matching conditions}
Relativistic hydrodynamics is formally constructed as an expansion around a state of local thermodynamic equilibrium, characterized by a temperature $T$, a chemical potential $\mu$, and a time-like four-velocity $u^\mu$. Away from equilibrium, however, these quantities do not admit a unique definition: different choices of $(T,\mu,u^\mu)$ can be used to parametrize the same conserved currents $T^{\mu\nu}$ and $N^\mu$.

To remove this ambiguity, one imposes additional constraints—known as matching conditions \cite{degroot1980,Denicol:2021}—which define how the hydrodynamic fields are identified in a non-equilibrium state. Equivalently, these conditions specify a choice of hydrodynamic frame, i.e., a prescription for decomposing the conserved currents into equilibrium and non-equilibrium contributions.

In this work we follow the Eckart picture \cite{Eckart:1940te}, in which the four-velocity is aligned with the net-charge 4-current, $N^\mu \equiv \mathcal{N} u^\mu$, so that the particle diffusion 4-current vanishes ($\mathcal{J^\mu} \equiv 0$).
The temperature and chemical potential are defined by requiring that the local energy and net-charge densities coincide with their equilibrium values as given by the equation of state, 
\begin{equation}
\mathcal{N} \equiv \mathrm{n}(T,\mu), \quad \mathcal{E} \equiv \mathrm{e}(T,\mu).    
\end{equation}
The isotropic pressure can then be decomposed into a thermodynamic pressure, determined by the equation of state, and a bulk viscous correction, 
\begin{equation}
\mathcal{P} = \mathrm{p}(T,\mu) + \Pi.
\end{equation}

\subsubsection{Israel-Stewart theory}

As discussed above, the conservation laws do not form a closed system and must be supplemented by additional relations specifying the form of the currents. A systematic way to construct such relations is to impose the second law of thermodynamics at the local level \cite{landau:59fluid},
\begin{equation}
\nabla_{\mu}S^{\mu }\geq 0,    
\end{equation}
where $S^\mu$ is the entropy 4-current. This approach provides a set of constraints on the allowed non-equilibrium contributions to the currents and, in particular, determines how dissipative effects enter the theory. 

A central difficulty in this construction is that the form of the entropy 4-current away from equilibrium is not uniquely determined from first principles. Nevertheless, in the vicinity of local equilibrium, the entropy current can be systematically organized as a series in the dissipative currents, regarded as small deviations from equilibrium, and truncated at a given order \cite{Israel:1979wp}. In this framework, first-order (Navier–Stokes) hydrodynamics arises from retaining terms linear in the dissipative quantities, while higher-order theories, such as that of Israel and Stewart, follow from including all contributions at quadratic order.

Then, to first order in the dissipative currents, the entropy 4-current can be expressed as \cite{landau:59fluid},
\begin{equation}
S^{\mu}=su^{\mu }+\frac{1}{T}Q^{\mu}, 
\end{equation}
with $s=s(T,\mu)$ being the entropy density in thermal equilibrium. The entropy production can then be
explicitly calculated from the conservation laws and has the following simple form \cite{Israel:1979wp},
\begin{equation}
\nabla _{\mu }S^{\mu }=-\frac{1}{T}\pi ^{\mu \nu }\nabla _{\mu }u_{\nu }-%
\frac{1}{T}\Pi \nabla _{\lambda }u^{\lambda } - Q^{\mu }\left( \dot{u}_{\mu }+%
\frac{\nabla _{\mu }T}{T}\right),
\end{equation}
where we introduce the notation $\dot{A} = u^{\alpha} \nabla_{\alpha}A$. The Navier–Stokes constitutive relations are then obtained as the most general first-order expressions, linear in the gradients, that render the entropy production non-negative for any fluid configuration \cite{landau:59fluid,Eckart:1940te},
\begin{subequations}
\label{eq:NS-cr}
\begin{align}
\pi ^{\mu \nu } &= -2\eta \sigma ^{\mu \nu }, \\
Q^{\mu } &= -\kappa \left( \dot{u}^{\mu }+
\frac{\nabla^{\mu }_{\perp}T}{T}\right) , \\
\Pi  &= -\zeta \nabla _{\lambda }u^{\lambda },
\end{align}%
\end{subequations}
where we defined the space-like gradient $\nabla_\perp^\mu \equiv \Delta^{\mu \nu} \nabla_\nu$ and introduced the transport coefficients $\eta ,\zeta ,\kappa >0$. The heat flow 4-current is thus intrinsically tied to temperature space-like gradients, while the viscous contributions are related to space-like gradients of velocity. 

While these constitutive relations are physically meaningful, they cannot be used as closure relations for the conservation laws. The resulting equations are parabolic and thus acausal, rendering the initial value problem ill posed \cite{hiscock:85generic,hiscock1983stability}. Thus, in relativistic settings, thermodynamic consistency and a gradient expansion alone are insufficient to guarantee a consistent theory, since the resulting equations must also define a well-posed and causal system of partial differential equations.

This issue can be addressed in more than one way; in the following we discuss the natural extension of the framework explained so far and describe how Israel and Stewart addressed this problem by considering second-order corrections to the entropy 4-current. In this
case, the entropy 4-current is corrected to \cite{Israel:1979wp},
\begin{equation}
S^{\mu }=su^{\mu }+\frac{1}{T}Q^{\mu }+\mathcal{O}_2^{\mu },    
\end{equation}
with $\mathcal{O}_2^{\mu }$ encoding all possible corrections to the entropy 4-current
that are of second-order in the dissipative currents,
\begin{eqnarray}
\mathcal{O}_2^{\mu }&=&-\left( \chi _{\Pi }\Pi ^{2}+\chi _{q}Q_{\alpha }Q^{\alpha
}+\chi_{\pi }\pi _{\alpha \beta }\pi ^{\alpha \beta  }\right) u^{\mu }   \nonumber \\ &+&  \chi
_{\Pi q}\Pi Q^{\mu }+\chi_{\pi q}\pi ^{\mu \nu }Q_{\nu }.
\end{eqnarray}
Above, the coefficients $\chi_i$ are general functions of temperature and chemical potential. These corrections modify the entropy production in such a way that imposing
\begin{equation}
\nabla _{\mu }S^{\mu }=\Pi ^{2}/(\zeta T)+\pi _{\alpha \beta }\pi ^{\alpha
\beta }/(2\eta T)+\mathcal{Q}_{\alpha }\mathcal{Q}^{\alpha }/\kappa \geq 0
\end{equation}
leads to novel equations of motion for the dissipative currents, the so-called Israel-Stewart theory \cite{Israel:1979wp},
\begin{subequations}
\begin{align}
\tau _{\Pi }\dot{\Pi}+\Pi  &= -\zeta \nabla _{\lambda }u^{\lambda }+\cdots. \label{eq:Bulk_IS}
\\
\tau _{q}\Delta _{\nu }^{\mu }\dot{Q}^{\nu }+Q^{\mu } &= 
 -\kappa \left( \dot{u}^{\mu }+
\frac{\nabla^{\mu }_{\perp}T}{T}\right) +\cdots , \label{eq:Heat_IS_accel}
\\
\tau _{\pi }\Delta _{\alpha \beta }^{\mu \nu }\dot{\pi}^{\alpha \beta }+\pi
^{\mu \nu } &= -2\eta \sigma ^{\mu \nu }+\cdots . \label{eq:Shear_IS}
\end{align}
\end{subequations}
Above, we introduced the relaxation times, $\tau _{\pi }=4\chi _{\pi
}T\eta $, $\tau _{q}=2\chi _{q}\kappa $ and $\tau _{\Pi }=2\chi _{\Pi
}T\zeta $, and the ellipsis denote higher-order terms that appear in this
derivation scheme --- since these terms will not be considered in this work, they were simply omitted. More complete second-order hydrodynamic equations can be derived from kinetic theory \cite{Denicol:2012cn,Denicol:2021,Rocha:2023ilf} or Holography \cite{BRSSS}.

The equation of motion for $Q^\mu$ can be rewritten in a more transparent form by eliminating the acceleration using energy-momentum conservation. Projecting $\nabla_\mu T^{\mu \nu}= 0$ orthogonally to $u^\mu$ yields
\begin{equation}
(\mathrm{e}+\mathrm{p})\dot{u}^\mu = -\nabla^{\mu}_{\perp}\mathrm{p} + \cdots,    
\end{equation}
which allows one to trade the acceleration for gradients of thermodynamic variables. Using standard thermodynamic identities, the heat-flow equation can then be recast as
\begin{equation}
\tau _{Q}\Delta _{\nu }^{\mu }\dot{Q}^{\nu }+Q^{\mu } =
 \lambda_Q \nabla^{\mu }_{\perp}\alpha +\cdots, \label{eq:Heat_IS}
\end{equation}
where $\alpha=\mu/T$ is the thermal potential and we introduced $\lambda_Q \equiv \kappa T\mathrm{n}/(\mathrm{e}+\mathrm{p})$. In this form, the driving force for heat flow is explicitly given by spatial gradients of an intensive thermodynamic quantity, as expected from irreversible thermodynamics \cite{Reichl2016}.
This change of variables also induces a redefinition of the relaxation time, $\tau_q \rightarrow \tau_Q$, as well as of the higher-order terms that have been omitted.

The Israel-Stewart equations \eqref{eq:Bulk_IS}, \eqref{eq:Shear_IS} and \eqref{eq:Heat_IS} can be rendered causal and linearly stable provided certain conditions are satisfied. In the linear regime, an analysis of small perturbations shows that it is sufficient to require\footnote{We shall briefly discuss the steps towards this constraint in Sec.~\ref{sec:perturbations-homog}. A more detailed and general analysis will be presented separately \cite{Guerrieri:inprep}.} 
\begin{align} \label{eq:causality_condition}
        &c_s^2 + \left[1+\frac{(\partial_{\mathrm{e}}\alpha)_{\mathrm{n}}}{(\partial_{\mathrm{n}} \alpha)_{\mathrm{e}}} (\partial_{\mathrm{n}} {\mathrm{p}})_\alpha \right] \frac{ \mathrm{n} (\partial_{\mathrm{n}}\alpha)_{\mathrm{e}}}{\mathrm{e} + \mathrm{p}} \frac{\lambda_Q}{\tau_Q} \nonumber \\
        &+ \frac{1}{\mathrm{e}+\mathrm{p}} \left[1+(\partial_{\mathrm{e}}\alpha)_{\mathrm{n}}\frac{\;\lambda_Q}{\tau_Q}\right] \left(\frac{4 \eta}{3\tau_\pi}+\frac{\zeta}{\tau_\Pi}\right) \leq 1,
    \end{align}
where $(\partial_x f)_y\equiv \partial f/\partial x|_y$ and $c_s$ is the adiabatic speed of sound, which may be expressed as
\begin{equation}
    c_s^2 \equiv (\partial_\mathrm{e} \mathrm{p})_\mathrm{s/n} = (\partial_\mathrm{e} \mathrm{p})_\mathrm{n} + \frac{\mathrm{n}}{\mathrm{e}+\mathrm{p}} (\partial_\mathrm{n} \mathrm{p})_\mathrm{e}.
\end{equation}
The inequality \eqref{eq:causality_condition} constrains the effective propagation speeds of the hydrodynamic modes to remain subluminal, with the terms proportional to $\zeta/\tau_\Pi$, $\eta/\tau_\pi$, and $\lambda_Q/\tau_q$ representing dissipative corrections to the characteristic speeds. In particular, this condition prevents taking the relaxation times to zero and recovering the Navier-Stokes regime: in that limit, dissipative signals propagate instantaneously and the equations become acausal and unstable \cite{hiscock:85generic,hiscock1983stability,Brito:2020nou,Denicol:2008ha,rischke,Sammet:2023bfo,Gavassino:2021kjm,Gavassino:2021owo}.

In this way, the fluid-dynamical framework proposed by Israel and Stewart may be understood as a thermodynamically constrained gradient expansion, providing a causal extension of relativistic hydrodynamics that incorporates the minimal set of additional structures required for consistency with the second law. It is also clear that in both Navier-Stokes and Israel-Stewart theories, a self-consistent treatment of heat conduction requires the existence of a finite temperature. In the following section, we discuss how we include a finite temperature in our treatment of a compact star.


\subsection{Equation of state}
\label{sec:eos}

In many studies of neutron-star structure and oscillations the stellar
matter is modeled as cold and described by a barotropic equation of state $
\mathrm{p}=\mathrm{p}(\mathrm{e})$. This approximation corresponds to cold catalyzed matter, where the
temperature is negligible compared with the microscopic energy scales of the
system, particularly the chemical potential, so that thermal contributions
to the pressure and other thermodynamic quantities can be ignored when
constructing the equilibrium configuration. While this simplification is
adequate for describing the background stellar structure, it cannot be
maintained in the presence of dissipative processes such as heat flow. Heat conduction is driven by temperature gradients and requires thermal excitations of the microscopic degrees of freedom, implying that the temperature cannot be strictly zero. The relevant physical regime in neutron
stars is therefore $T\ll \mu $ rather than $T=0$. Consequently, a
consistent hydrodynamic description of a dissipative stellar fluid must
include temperature (or equivalently the entropy) as an independent
thermodynamic variable, even if thermal effects remain subdominant in determining the equilibrium equation of state. In this section, we detail our choice of EoS.

Given the uncertainties in modeling the nuclear EoS, parametrized models are commonly used in the literature. The simplest of these is likely the polytropic form, $\mathrm{p}(\mathrm{n}) = K \mathrm{n}^\gamma$, where the polytropic exponent $\gamma$ controls the stiffness of the EoS. As a one-parameter relation, this prescription does not completely fix the thermodynamic state unless additional information -- such as the temperature, internal energy, or entropy -- is specified. It is often used under the assumption of completely degenerate matter ($T=0$), in which case the first law of thermodynamics implies $\mathrm{e}(\mathrm{n}) = m_b \mathrm{n} + \mathrm{p}(\mathrm{n})/(\gamma -1)$. These same relations for $\mathrm{p}(\mathrm{n})$ and $\mathrm{e}(\mathrm{n})$ also emerge in the different context of an isentropic ideal gas \cite{Rezzolla:2013dea}. 

In recent studies of viscous stellar dynamics, e.g.~\cite{Shum:2025jnl,Keeble:2026bzo}, a polytropic EoS has been employed as a convenient closure relation even in the presence of heat flux, within the BDNK framework. However, as noted previously, when heat conduction is included, the pressure and energy density can no longer be specified by a barotropic relation alone, and an EoS that tracks temperature or entropy as an independent thermodynamic degree of freedom becomes necessary. A step in this direction was taken in Ref.~\cite{Pandya:2022sff}, where BDNK was supplemented with ideal-gas microphysics including a finite-temperature equation of state. Nevertheless, for neutron stars, where thermal contributions to the pressure are subdominant and the equilibrium structure is governed by cold, dense nuclear matter, an ideal-gas EoS remains of limited applicability. 

An alternative, physically motivated approach is the hybrid decomposition often adopted in the numerical relativity literature (see e.g.~\cite{Figura:2020fkj,guerra2025treatmentthermaleffectsequation}), in which the EoS is given by a sum of cold and thermal contributions. The cold sector may be modeled by a simple polytrope or by a more realistic parametrization, such as a piecewise-polytropic fit \cite{Read:2008iy,OBoyle:2020qvf}, while the thermal sector is often described by an ideal-gas prescription. We draw inspiration from this approach, and also from the Sommerfeld expansion \cite{Greiner:1995vqa} of the EoS for a degenerate Fermi gas to propose a simple, parametrized model for the EoS that (i) is thermodynamically consistent and (ii) includes finite-temperature effects as controlled corrections, yielding parametrically small thermal contributions to the equilibrium configurations (even in the low-density regime near the stellar surface).

Specifically, we adopt the following parametrization for the net-baryon density, pressure, and energy density:
\begin{widetext}
\begin{subequations}
\label{eq:n_eos}
    \begin{align} 
    \label{eq:n_eos_n}
    \mathrm{n}(\hat{\mu},T) & = A \hat{\mu}^\lambda \left(1+ w \frac{T^2}{\mu^2}\right),\\
    \label{eq:n_eos_p}
    \mathrm{p}(\hat{\mu},T) & = \frac{A \hat{\mu}^{\lambda+1}}{\lambda +1} \left[1 + w \frac{T^2}{m_b^2} {}_2F_1(2,1+\lambda;2+\lambda;-\hat{\mu}/m_b)
    \right],\\
    \label{eq:n_eos_e}
    \mathrm{e}(\hat{\mu},T)  & = m_b \mathrm{n}(\hat{\mu},T) + \frac{\lambda}{\lambda +1} A \hat{\mu}^{\lambda+1} + A w \hat{\mu}^{\lambda+1} \frac{T^2}{\mu^2} \left[ 1 + \frac{\mu^2}{m_b^2} \frac{1}{1+\lambda} {}_2F_1(2,1+\lambda;2+\lambda;-\hat{\mu}/m_b)
    \right ].
\end{align}
\end{subequations}
\end{widetext}
Here $A$, $w$ and $\lambda\equiv 1/(\gamma-1)$ are constant parameters, ${}_2F_1(a,b;c;z)$ is the  hypergeometric function \cite{gradshteyn2014table}, and $$\hat{\mu} \equiv \mu - m_b$$ is the mass-subtracted chemical potential, with $m_b$ denoting the neutron mass. At $T=0$, one recovers a simple polytrope, written in the standard form $\mathrm{p}(\mathrm{n})=K\mathrm{n}^\gamma$ upon the redefinition $K = A^{-1/\lambda}(1+\lambda)^{-1}$. 

The expression for $\mathrm{n}$, Eq.~\eqref{eq:n_eos_n}, possesses a cold and a thermal piece, in a form that resembles the low-temperature limit (also known as the Sommerfeld expansion \cite{Greiner:1995vqa}) of the EoS for a non-relativistic degenerate Fermi gas, in which case $\lambda = 3/2$. A key difference, though, is that the Sommerfeld expansion is naturally organized in powers of $(T/\hat{\mu})$, whereas here the thermal corrections in Eq.~\eqref{eq:n_eos_n} scale as $(T/\mu)^2 = [T/(\hat{\mu}+m)]^2$. Our main motivation for doing so is to keep temperature corrections under control even when the EoS is applied up to the stellar crust (where $\hat{\mu} \to 0$). For high chemical potentials, one recovers $\mu \approx \hat{\mu}$, and the standard Sommerfeld scaling is effectively recovered.

From $\mathrm{n}(\hat{\mu},T)$ in Eq.~\eqref{eq:n_eos_n}, expression \eqref{eq:n_eos_p} for the pressure follows from integrating the Maxwell relation $\mathrm{n} = (\partial \mathrm{p}/\partial \hat{\mu})_T$. Similarly, the entropy density can be obtained by integrating $s = (\partial \mathrm{p}/\partial T)_\mu$:
\begin{equation}
    s(\hat{\mu},T) = \frac{A \hat{\mu}^{\lambda+1}}{\lambda +1} \frac{2 w T}{m_b^2} {}_2F_1(2,1+\lambda;2+\lambda;-\hat{\mu}/m_b).
\end{equation}
Finally, the energy density in Eq.~\eqref{eq:n_eos_e} follows from the Euler relation $\mathrm{e}+\mathrm{p} = Ts + \mu \mathrm{n}$. The system \eqref{eq:n_eos} is therefore thermodynamically consistent by construction. 

\begin{figure}[tbh] 
    \includegraphics[width=0.98\linewidth]{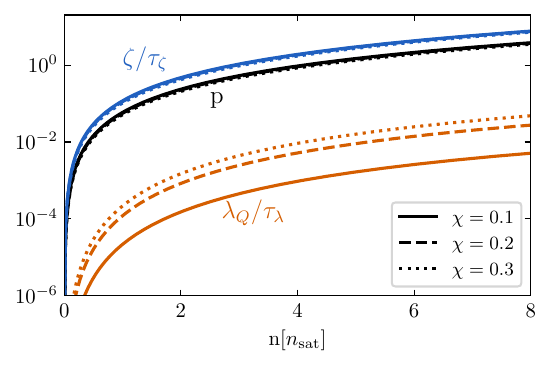}
    \caption{Density profiles of the thermodynamic pressure ${\rm p}$ (black) and of the transport combinations $\zeta/\tau_\zeta$ (blue) and $\lambda_Q/\tau_\lambda$ (orange), all in units of $\rho_{\rm sat}c^2$, for the EoS and transport prescriptions employed in this work, and assuming $T = \chi \mu$, with $\chi \in \{0.1,0.2, 0.3\}$.}
    \label{fig:EoS_transport}
\end{figure}

For applications in the present paper, we adopt $\lambda = 1$ ($\gamma =2$) and $A = 25 \rho_{\rm sat} / (3 m_b^2)$ ($\rho_\mathrm{sat} = m_b n_{\rm sat} = 2.7 \times 10^{14} \textrm{g/cm}^3$), which, for $T=0$, matches the EoS used in Paper I \cite{Mendes:2025oib}. 
Figure \ref{fig:EoS_transport} displays, in black, the thermodynamic pressure as a function of number density along the equilibrium line $T=\chi \mu$ (corresponding to diffusive equilibrium; see Eq.~\eqref{eq:diffusive_equilibrium} below), for several values of $\chi$. The pressure exhibits only a weak dependence on temperature, with $O(\chi^2)$ thermal corrections remaining modest across the entire density range considered.

\subsection{Transport coefficients} \label{sec:transport}

A variety of microscopic processes contribute to viscous dissipation and heat transport in neutron-star matter. As a result, transport coefficients such as the bulk viscosity \cite{Sawyer:1989dp,Haensel:1992zz,Yang:2023ogo}, shear viscosity \cite{Shternin:2008es,Manuel:2012rd}, and thermal conductivity \cite{Shternin:2007ee,Goodwin:1982hy} can, in principle, be computed from an underlying microscopic description of the system. In the present work, however, we adopt a more phenomenological approach to specifying the transport coefficients entering the Israel–Stewart equations. Rather than relying on detailed microphysical calculations, we choose functional forms guided by mathematical simplicity, and constrained by general physical requirements, in particular the causality condition \eqref{eq:causality_condition}.

For the bulk and shear viscosities, we set 
\begin{equation} \label{eq:transport_eta_zeta}
    \zeta = \tau_\zeta \mathrm{n} (\partial_\mathrm{n} \mathrm{p})_\alpha, \qquad 
    \eta = \tau_\eta \mathrm{n} (\partial_\mathrm{n} \mathrm{p})_\alpha.
\end{equation}
As it will be discussed in Sec.~\ref{sec:radial_perturbations}, this choice, together with Eq.~\eqref{eq:tau_prescription} below for the $\tau_i$'s, avoids the appearance of singular points in the integration of the frequency-domain equations. Moreover, for a polytropic EoS at $T=0$, one has $\mathrm{n} (\partial_\mathrm{n} \mathrm{p})_\alpha = (\mathrm{e}+\mathrm{p})c_s^2$. In this limit, and in the absence of heat diffusion ($\lambda_Q = 0$), the causality condition \eqref{eq:causality_condition} can be expressed compactly as
\begin{equation}
c_s^2 \left( 1 + \frac{4\tau_\eta}{3\tau_\pi} + \frac{\tau_\zeta}{\tau_\Pi} \right) \leq 1,
\end{equation}
which constrains the dimensionless ratios $\tau_\eta/\tau_\pi$ and $\tau_\zeta/\tau_\Pi$. For the equation of state adopted here, small but finite temperatures introduce only minor corrections to this relation.

For the heat transport coefficient, we assume 
\begin{equation} \label{eq:lambdaQ_prescription}
    \lambda_Q = \tau_\lambda \frac{c_s^2  m_b}{(\partial_\mathrm{n}\alpha)_\mathrm{e}}.
\end{equation}
For the EoS employed in this work, one has $(\partial_\mathrm{n}\alpha)_\mathrm{e} \sim T^{-3}$ as $T \to 0$, implying $\lambda_Q \sim T^3$ in this limit. The choice \eqref{eq:lambdaQ_prescription} is primarily motivated by the causality condition \eqref{eq:causality_condition}. In particular, if $\lambda_Q$ were to vanish more slowly than $T^3$ as $T \to 0$, the causality condition would be violated in this limit. With the prescription \eqref{eq:lambdaQ_prescription}, and in the limit $\eta = \zeta = 0$, Eq.~\eqref{eq:causality_condition} can be written as
\begin{equation} \label{eq:causality_restricted}
    c_s^2\left(1 + b \frac{\tau_\lambda}{\tau_Q} \right) \leq 1,
\end{equation}
where $b = \left[1 + (\partial_{\mathrm{e}}\alpha)_{\mathrm{n}} (\partial_{\mathrm{n}} {\mathrm{p}})_\alpha/(\partial_{\mathrm{n}} \alpha)_{\mathrm{e}}\right] (m_b\mathrm{n})/(\mathrm{e} + \mathrm{p}) \sim 1$ is a coefficient that varies only weakly with density and temperature. 

The transport combinations $\zeta/\tau_\zeta (=\eta/\tau_\eta)$ and $\lambda_Q/\tau_\lambda$ are shown in Fig.~\ref{fig:EoS_transport} in blue and orange, respectively, along the diffusive-equilibrium line $T = \chi \mu$  [cf.~Eq.~\eqref{eq:diffusive_equilibrium} below] for a few values of $\chi$, under the assumption of the EoS described in Sec.~\ref{sec:eos}. The former depends only weakly on temperature, whereas the latter has a more pronounced temperature dependence. 

We stress the distinct physical meanings of the relaxation times ${\tau_Q,\tau_\Pi,\tau_\pi}$ and the time-dimensioned parameters ${\tau_\lambda,\tau_\zeta,\tau_\eta}$: The latter determine the magnitudes of the transport coefficients $\lambda_Q$, $\zeta$, and $\eta$ through Eqs.~(20) and (22).
Finally, for $i \in \{\zeta,\eta,\lambda,\Pi,\pi,Q\}$ we set
\begin{equation} \label{eq:tau_prescription}
    \tau_i = e^{\Phi/2} t_i, \qquad t_i = \textrm{cte}.
\end{equation}
The redshift factor $e^{\Phi/2}$ is an $O(1)$ quantity introduced, as in Paper I \cite{Mendes:2025oib}, to avoid the appearance of singular points in the integration domain of the frequency-domain equations.
Note that the ratios ${\tau_\eta/\tau_\pi,\tau_\zeta/\tau_\Pi,\tau_\lambda/\tau_Q}$ entering the causality relations (21) and (23) are equivalent to the corresponding ratios ${t_\eta/t_\pi,t_\zeta/t_\Pi,t_\lambda/t_Q}$ of the redshifted times introduced in Eq.~(24).


\section{Perturbative analyses} \label{sec:perturbations}

\subsection{Perturbations around global equilibrium in Minkowski spacetime}
\label{sec:perturbations-homog}

A useful perspective for interpreting the oscillation spectrum of viscous stars is provided by the dispersion relations of perturbations around homogeneous equilibrium states. In an infinite medium, translational invariance allows perturbations to be decomposed into plane waves labeled by a continuous wave number $k$, leading to dispersion relations of the form $\omega=\omega(k)$. In contrast, the oscillation spectrum of a finite star consists of a discrete set of eigenfrequencies \{$\omega_n$\}, determined by regularity conditions at the stellar center together with boundary conditions at the stellar surface \cite{Kokkotas:1999bd,Nollert:1999ji}.

At first sight, these two problems appear rather different. A relativistic star is finite, strongly inhomogeneous, and dynamically coupled to gravity, so one should not expect a quantitative identification between the continuous hydrodynamic spectrum and the discrete stellar spectrum. Nevertheless, it was recently observed in Paper I \cite{Mendes:2025oib} that many qualitative features of the dispersion relations of homogeneous relativistic hydrodynamics survive in the spectrum of radial oscillations of viscous stars. In particular, the stellar overtones were shown to organize themselves in a manner closely analogous to the hydrodynamic and non-hydrodynamic branches appearing in the homogeneous problem. This observation suggests that the stellar spectrum may be understood, at least qualitatively, as a finite-size realization of the underlying hydrodynamic mode structure, with the finite radius of the star effectively discretizing the continuous dispersion relations. 


Motivated by these considerations, we now review the dispersion relations associated with perturbations around homogeneous equilibrium configurations in the presence of heat flow. An initial equilibrium state, characterized by the hydrodynamic fields 
$\{\mathrm{n}, \mathrm{e}, u^\mu\}$ is perturbed to a nearby non-equilibrium configuration, while the background spacetime metric is assumed to remain flat and fixed. From the linearized local conservation laws \eqref{eq:CL} and the Israel-Stewart equations of motion, \eqref{eq:Bulk_IS}, \eqref{eq:Shear_IS} and \eqref{eq:Heat_IS}, one obtains the dispersion relations governing the perturbations in Fourier space. As in Paper I \cite{Mendes:2025oib}, we focus on longitudinal perturbations, which encode the dissipative corrections to sound propagation. The corresponding dispersion relations are obtained as the roots of the following sixth-order polynomial in $\omega(k)$ \cite{Guerrieri:inprep}:
\begin{align} \label{eq:determinant}
   0&= i\omega (1- i \omega \tau_Q)  \bigg\{\frac{i \omega k^2}{ \mathrm{e} + \mathrm{p}} \left[\zeta (1- i \omega\tau_\pi) + \frac{4}{3} \eta (1- i \omega \tau_\Pi)\right] \nonumber \\
   & + (\omega^2 - k^2 c_s^2) (1- i \omega \tau_\Pi)(1- i \omega\tau_\pi)\bigg\}\nonumber\\
    &+\frac{\lambda_{Q}  k^2}{\mathrm{e} + \mathrm{p}} \bigg\{i\omega k^2 (\partial_\mathrm{e}\alpha)_n  \left[ \zeta (1- i \omega\tau_\pi) + \frac{4}{3} \eta (1- i \omega \tau_\Pi)\right] \nonumber \\
    & - n (\partial_\mathrm{n}\alpha)_\mathrm{e} \left[ \omega^2 -(\partial_\mathrm{e}p)_\alpha k^2 \right](1- i \omega \tau_\Pi)(1- i \omega\tau_\pi)\bigg\},
\end{align}
where $k$ denotes the magnitude of the wave vector. Figure \ref{fig:intro_modes} depicts these dispersion relations for representative parameter choices in the cases of (a) bulk (or shear) viscosity only and (b) heat diffusion only.

In particular, we may draw intuition on the propagation of these modes from the small $k$ (large wavelength) limit. Up to $O(k^2)$, the six solutions take the following form
\begin{subequations}\label{eq:Taylor_omega}
\begin{align}
    \omega^\pm_s (k) & = \pm c_s k - i \Gamma_s k^2 + O(k^3), \\
    \omega_j(k) & = -i \left(\frac{1}{\tau_j} - \Gamma_j k^2 \right) + O(k^3), \quad j \in \{\pi,\Pi,Q\}, \\
    \omega_T (k) & = i k^2 \left( 2 \Gamma_s - \Gamma_\pi - \Gamma_\Pi - \Gamma_Q\right) + O(k^3),
\end{align}
\end{subequations}
where we have introduced
\begin{equation}
    \Gamma_\pi \equiv \frac{4\eta}{3(\mathrm{e} + \mathrm{p})}, \quad
    \Gamma_\Pi \equiv \frac{\zeta}{\mathrm{e} + \mathrm{p}}, \quad
    \Gamma_Q \equiv \frac{\lambda_Q \mathrm{n}(\partial_\mathrm{n}\alpha)_\mathrm{e}}{\mathrm{e} + \mathrm{p}},
\end{equation}
and
\begin{equation}
    2 \Gamma_s \equiv \Gamma_\pi + \Gamma_\Pi + \Gamma_Q \frac{[(\partial_\mathrm{n}\mathrm{p})_\mathrm{e}]^2}{T (\partial_\mathrm{n}\alpha)_\mathrm{e} c_s^2(\mathrm{e}+\mathrm{p})}.
\end{equation}

The spectrum therefore consists of two propagating sound modes ($\omega_s^\pm$), a thermal mode ($\omega_T$) which is purely diffusive at small $k$, and three non-hydrodynamic modes ($\omega_\pi, \omega_\Pi, \omega_Q$) whose frequencies remain finite as $k \to 0$. Among the latter, $\omega_Q$ is associated with heat-flux relaxation. At small $k$, both $\omega_T$ and $\omega_Q$ are purely damped. At a critical wave number, $k_c$, they collide on the imaginary axis. For $k > k_c$, they acquire equal and opposite real parts while sharing the same imaginary part, forming a pair of propagating thermal waves -- the relativistic analogue of second sound \cite{Joseph:1989zz,Andersson:2011fd}. In the opposite wavelength limit, $k\to \infty$, requiring that the characteristic group velocities of all propagating modes remain subluminal leads to the causality condition \eqref{eq:causality_condition} \cite{Guerrieri:inprep}. 

In the stellar problem, one expects these continuum branches to manifest themselves as discrete families of modes labeled by the overtone number $n$, with increasing $n$ probing progressively shorter effective wavelengths within the star. We will return to this interpretation in Sec.~\ref{sec:results}.

\subsection{Radial perturbations around a static, spherically symmetric star} \label{sec:radial_perturbations}

In this section, we specialize the equations of motion \eqref{eq:CL} and field equations \eqref{eq:EFE} to spherically symmetric systems. For brevity, we outline only the main steps of the derivation; a more detailed discussion can be found in Paper I \cite{Mendes:2025oib} (see also Refs.~\cite{Redondo-Yuste:2024vdb,Keeble:2026bzo}).

In appropriate coordinates, the spacetime can be described by the line element
\begin{equation}
    ds^2 = - e^{\Phi (t,r)} dt^2 + e^{\Lambda(t,r)} dr^2 + r^2 (d\theta^2 + \sin^2\theta d\phi^2),
\end{equation}
where the metric potentials are decomposed into a static background contribution and a time-dependent perturbation:
\begin{subequations}
\begin{align}
    \Phi(t,r) & = \Phi_{0}(r) + \delta \Phi (t,r), \\
    \Lambda(t,r) & = \Lambda_{0}(r) + \delta \Lambda (t,r).
\end{align}
\end{subequations}

In spherical symmetry, it is convenient to introduce a local basis on the two-dimensional subspace orthogonal to the symmetry 2-spheres, adapted to the fluid flow. This basis is formed by the fluid four-velocity $u^\mu$ and a spacelike unit vector field $m^\mu$, satisfying $u_\mu m^\mu = 0$. To linear order in the perturbations, these vector fields take the form
\begin{subequations}
    \label{eq:pert-u}
\begin{align}
    u^\mu & = e^{-\Phi_{0}/2} \left( 1- \delta \Phi/2, \dot{\xi},0,0 \right),  \\
    m^\mu & = e^{-\Lambda_{0}/2} ( e^{\Lambda_0-\Phi_0} \dot{\xi}, 1 - \delta \Lambda/2,0,0 ),
\end{align}    
\end{subequations}
where $\xi(t,r)$ denotes the Lagrangian displacement.
In terms of this local basis, the energy-diffusion four-current, the shear stress tensor, the net-baryon four-current and the energy-momentum tensor can be expressed as
\begin{subequations}
\begin{align}
    Q^\mu &= Q(t,r) m^\mu, 
    \\
    \pi_{\mu\nu} & = \tilde{\pi}(t,r) m_\mu m_\nu - \frac{1}{2} \tilde{\pi} r^2 \gamma_{\mu\nu} ,
    \\
    N^\mu &= \mathrm{n}(t,r) u^\mu , 
    \\
    T_{\mu\nu} &= 
    \mathcal{E}(t,r) u_\mu u_\nu + [\mathcal{P}(t,r) + \tilde{\pi}(t,r)] m_\mu m_\nu \nonumber \\
    & + 2 Q(t,r) m_{(\mu} u_{\nu)}  + r^2 \gamma_{\mu\nu} [\mathcal{P}(t,r) -\tilde{\pi}(t,r)/2] ,
\end{align}
\end{subequations}
where $\gamma_{\mu\nu} = \textrm{diag}(0,0,1,\sin^2\theta)$ is the metric of the unit 2-sphere.
We consider linear perturbations of fluid-related variables so that
\begin{subequations}
\begin{align}
\mathrm{n}(t,r) &= \mathrm{n}_{0}(r) + \delta \mathrm{n}(t,r), \\
 \mathrm{e}(t,r) &= \mathrm{e}_{0}(r) + \delta \mathrm{e}(t,r), \\
    \mathcal{P}(t,r) &= \mathrm{p}_{0}(r) + \delta \textrm{p} (t,r) + \Pi (t,r). 
\end{align}
\end{subequations}

The background quantities are related to each other through the equation of state $\mathrm{p}_{0}(\mathrm{n}_{0}, \mathrm{e}_{0})$ and the Tolman-Oppenheimer-Volkoff (TOV) equations for a perfect fluid:
\begin{align}
    \Lambda_0' & = \frac{e^{\Lambda_0}}{r} (e^{-{\Lambda_0}}-1 + 8\pi r^2 \textrm{e}_0), \\
    \Phi_0' & = \frac{e^{\Lambda_0}}{r} (1-e^{-\Lambda_0} + 8 \pi r^2 \textrm{p}_0), \\
    \textrm{p}_0' & = -(\textrm{e}_0 + \textrm{p}_0) \frac{\Phi_0'}{2},
\end{align}
with $f' \equiv df/dr$. The TOV equations can be numerically integrated from $r = 0$ to the stellar radius $R$, defined by $\mathrm{p}_0(R) = 0$. The total mass $M$ can then be obtained as $M = R [1 - e^{-\Lambda_0(R)}]$, and the stellar compactness is the dimensionless ratio $C = M/R$.

For a two-parameter EoS, the TOV equations provide three equations for four unknowns $\{\Lambda_0$, $\Phi_0$, $\mathrm{e}_0$, $\mathrm{n}_0 \}$, with $\mathrm{p}_0 = \mathrm{p}_0(\mathrm{n}_0,\mathrm{e}_0)$, so one additional condition is required to close the system. We use this freedom to require that the heat current vanishes at the background level, so that the dissipative current variables $\tilde{\pi}(t,r)$, $\Pi(t,r)$, and $Q(t,r)$ may all be consistently treated as first-order perturbations. From Eq.~\eqref{eq:Heat_IS} the condition of diffusive equilibrium implies that
\begin{equation} \label{eq:diffusive_equilibrium}
    \alpha_0' =(\mu_0/T_0)' = 0 \quad \Rightarrow \quad T_0(r) = \chi \mu_0(r),
\end{equation}
where $\chi$ is a non-negative constant.

The linearized equations of motion for the perturbation variables may be organized into three groups (see Paper I \cite{Mendes:2025oib}), where the `0' subindex for background variables shall be omitted for notation simplicity. The first arises from the Einstein field equations \eqref{eq:EFE},  
\begin{subequations}
\label{eq:EFE-pert}
\begin{align}
\label{eq:dLambdadt}
    \dot{\delta \Lambda} & = - 8\pi r e^{\Lambda} [ (\textrm{e} + \textrm{p}) \dot{\xi} + e^{(\Phi-\Lambda)/2} Q],
\\ 
\delta \Phi' &= 8\pi r e^{\Lambda} (\delta \textrm{p} + \Pi + \tilde{\pi} ) + \frac{e^{\Lambda}}{r}  (1 + 8 \pi r^2 \textrm{p}) \delta \Lambda.
\end{align}    
\end{subequations}
The second stems from the local conservation laws for energy-momentum and net-baryon four-current \eqref{eq:CL-emt} and \eqref{eq:CL-d4c}, 
\begin{subequations}
\label{eq:consv-pert}
\begin{align}
\label{eq:deltan}
    \frac{\delta \textrm{n}}{\textrm{n}} + \frac{1}{2} \delta \Lambda + \xi \left( \frac{2}{r} + \frac{1}{2} \Lambda' + \frac{\textrm{n}'}{\textrm{n}} \right) + \xi' &= 0,\\
\label{eq:dedt}
    \dot{\delta \textrm{e}} + \frac{1}{2} (\textrm{e} + \textrm{p}) \dot{\delta\Lambda} + \left( \textrm{e}' + \frac{1}{2r} (\textrm{e} + \textrm{p})(4 + r \Lambda') \right) \dot{\xi} 
    \notag
    & \\
    + (\textrm{e} + \textrm{p}) \dot{\xi'} +e^{(\Phi-\Lambda)/2} \left[ Q' + Q \left( \frac{2}{r} + \Phi' \right)\right] &= 0,\\
    e^{\Lambda - \Phi} (\textrm{e} + \textrm{p}) \ddot{\xi} + \frac{1}{2} (\textrm{e} + \textrm{p}) \delta \Phi' + \delta \textrm{p}' + \Pi' + \tilde{\pi}'
     & \notag
    \\
    + \frac{1}{2} \Phi' (\delta \textrm{p} + \delta \textrm{e} + \Pi + \tilde{\pi} )
    + \frac{3}{r} \tilde{\pi}
    + e^{(\Lambda-\Phi)/2} \dot{Q} & = 0. 
\end{align}
\end{subequations} 
The third set of equation stems from the Israel-Stewart equations of motion \eqref{eq:Bulk_IS}, \eqref{eq:Heat_IS} and \eqref{eq:Shear_IS}, respectively, 
\begin{subequations}
\label{eq:IS-NS-eoms}
\begin{align}
\tau_\Pi \dot{\Pi} + e^{\Phi/2} \Pi = & - \zeta \left[ \dot{\xi}' + \frac{1}{2}\dot{\delta \Lambda} + \dot{\xi} \left( \frac{2}{r} +\frac{\Lambda'}{2} \right)
    \right], \\
\label{eq:IS-NS-eoms-Q}    
    \tau_Q \dot{Q} + e^{\Phi/2} Q = & e^{(\Phi-\Lambda)/2} \lambda_Q \delta \alpha',
    \\
    \tau_\pi \dot{\tilde{\pi}} + e^{\Phi/2} \tilde{\pi} = & - \frac{4}{3} \eta \left[ \dot{\xi}' + \frac{1}{2}\dot{\delta \Lambda} + \dot{\xi} \left(- \frac{1}{r} +\frac{\Lambda'}{2} \right)
    \right].    
\end{align}    
\end{subequations}
Additionally, the equation of state may be used to connect $\delta \mathrm{p}$ and $\delta \alpha$ to the primary variables $\delta \mathrm{n}$ and $\delta \mathrm{e}$:
\begin{align}
\delta \mathrm{p} & = (\partial_\mathrm{n} \mathrm{p})_\mathrm{e} \delta \mathrm{n} + (\partial_\mathrm{e} \mathrm{p})_\mathrm{n} \delta \textrm{e}, \\
\delta \alpha & = (\partial_\mathrm{n} \alpha)_\mathrm{e} \delta \mathrm{n} + (\partial_\mathrm{e} \alpha)_\mathrm{n} \delta \textrm{e},
\end{align}
where partial derivatives can be conveniently computed from Eq.~\eqref{eq:n_eos} -- formulated in terms of $(\hat{\mu},T)$ variables -- with the aid of Jacobian identities such as
\begin{equation}
    (\partial_\mathrm{n} \mathrm{p})_\mathrm{e} = \frac{(\partial_{\hat \mu} \mathrm{p})_T (\partial_T \mathrm{e})_{\hat \mu} - (\partial_T \mathrm{p})_{\hat \mu} (\partial_{\hat \mu} \mathrm{e})_T}{(\partial_{\hat \mu} \mathrm{n})_T (\partial_T \mathrm{e})_{\hat \mu} - (\partial_T \mathrm{n})_{\hat \mu} (\partial_{\hat \mu} \mathrm{e})_T}.
\end{equation}

Equations \eqref{eq:EFE-pert}, \eqref{eq:consv-pert} and \eqref{eq:IS-NS-eoms} form a system of linear partial differential equations for the   $(t,r)$-dependent functions $\{ \xi, \delta \Lambda, \delta \Phi, \delta \textrm{n}, \delta \textrm{e}, \Pi, Q, \tilde{\pi} \}$, with $r$-dependent coefficients $\{\textrm{n}, \textrm{e}, \Phi, \Lambda\}$, related to the background fluid. Moreover,  $\{\textrm{p}, \zeta, \lambda_Q, \eta, \tau_\Pi, \tau_Q, \tau_\pi\}$ are functions of $(\textrm{n},\textrm{e})$ (or, equivalently, $(\mu, T)$ or $(\hat{\mu}, T)$), which codify both transport and equation of state properties. 

Assuming a harmonic dependence
\begin{equation}
    y(t,r) = e^{-i\omega t} y(r) ,       
\end{equation}
for all perturbation variables, the system formed by Eqs.~\eqref{eq:EFE-pert}, \eqref{eq:consv-pert}, and \eqref{eq:IS-NS-eoms} can be recast as a linear, homogeneous, second-order master equation for the state vector ${\bf X} = (\xi,Q)^T$, schematically given by
\begin{align} \label{eq:pert}
    a {\bf X}'' + {\bf B} \cdot {\bf X}' + {\bf C} \cdot {\bf X} = 0,
\end{align}
where ${\bf B}(r)$ and ${\bf C}(r)$ are coefficient matrices depending solely on background functions, and 
\begin{equation}
    a(r) = \textrm{n} (\partial_\textrm{n} \textrm{p})_\alpha + \frac{\omega \zeta}{i e^{\Phi/2} + \omega \tau_\Pi} + \frac{4}{3} \frac{\omega \eta}{i e^{\Phi/2}+\omega \tau_\pi},
\end{equation}
where we notice that
\begin{equation}
    (\partial_\textrm{n} \textrm{p})_\alpha = (\partial_\textrm{n} \textrm{p})_\textrm{e} - \frac{ (\partial_\textrm{e} \textrm{p})_\textrm{n} (\partial_\textrm{n} \alpha)_\textrm{e} }{(\partial_\textrm{e} \alpha)_\textrm{n}}.
\end{equation}
Points at which $a(r)=0$  are singular, as the equations cease to determine the second derivatives uniquely. The choices \eqref{eq:transport_eta_zeta} and \eqref{eq:tau_prescription}, however, prevent such singular points from appearing within the integration domain, and are therefore technically convenient. Further discussion may be found in Paper I.

Solutions of Eq.~\eqref{eq:pert} that are regular at $r=0$ must satisfy 
\begin{equation} \label{eq:bc_origin}
    {\bf X}(0) = (0,0),
\end{equation}
while ${\bf X}'(0) = (\xi_1,Q_1)$ is left unspecified by regularity alone. At the stellar surface ($r = R$), the heat current must vanish, since there is no medium beyond the star to support diffusion. For our choices of equation of state and transport coefficients, 
\begin{equation} \label{eq:bc_R}
    Q(r) \approx Q_R \, (r - R)^{\lambda + 2},
\end{equation}
as $r \to R$, where $\lambda$ is related to the polytropic exponent and $Q_R$ is a free amplitude. Imposing regularity at the stellar surface --- which is sufficient to ensure that the Lagrangian pressure perturbation vanishes there --- yields a relation for $\xi'(R)$ in terms of $\xi_R \equiv \xi(R)$ and $Q_R$.
The problem thus involves two free parameters at each boundary [$(\xi_1, Q_1)$ at the center and $(\xi_R, Q_R)$ at the surface], but an overall normalization eliminates one of these. Integrating the system \eqref{eq:pert} from $r=0$ to some intermediate $r_\mathrm{match}$, and from $R$ to $r_\mathrm{match}$, subject to the boundary conditions above, leave four matching conditions at $r_\mathrm{match}$ for the inner and outer solutions to join together smoothly. These cannot be satisfied by adjusting the three remaining coefficients for generic $\omega$, but only for a discrete set of complex eigenvalues $\omega_n$, where $n \in \mathbb{N}$. 

The coefficient matrices ${\bf B}(r)$ and ${\bf C}(r)$ in Eq.~\eqref{eq:pert} are too lengthy to be presented here, but some of their properties can be highlighted. In particular, for our choice of EoS and transport coefficients, in the zero-temperature limit these coefficient matrices acquire a triangular structure: the equation for $Q$ decouples from $\xi$, while the equation for $\xi$ retains a source term depending on $Q$ and $Q'$. As a consequence, the spectrum of the system decomposes into two sectors. The first consists of eigenvalues $\omega_n^{(\xi)}$ obtained by setting $Q = 0$; these include the standard fluid oscillation modes, modified by bulk and viscous corrections, as well as the non-hydrodynamic modes associated with bulk and shear viscosity, none of which are affected by the heat flux in this limit.
The second consists of eigenvalues $\omega_n^{(Q)}$ determined by the decoupled heat equation alone. At each such frequency, the $\xi$ equation becomes an inhomogeneous ODE with the corresponding $Q$ eigenfunction as a fixed source; this equation can generically be solved  (unless, potentially, in the non-generic case where $\omega_n^{(Q)}$ coincides with an eigenvalue $\omega_m^{(\xi)}$). At finite temperature, the triangular structure is broken and the two sectors mix, but the decoupled spectra $\{\omega_n^{(\xi)}\} \cup \{\omega_n^{(Q)}\}$ provide natural starting points for tracking eigenvalues as $T$ is increased from zero.

\section{Numerical results} \label{sec:results}

In this section, we present results for the complex eigenvalues of the system \eqref{eq:pert}, subject to the boundary conditions \eqref{eq:bc_origin} and \eqref{eq:bc_R}. The equations are numerically integrated outward from $r\approx 0$ to a matching radius $r_{\rm match}\approx R/2$, and inward from $r\approx R$ to the same point. Requiring continuity of ${\bf X}$ and ${\bf X}'$ at $r_{\rm match}$ yields four matching conditions. For a given trial frequency $\omega\in\mathbb{C}$, three of these conditions are used to determine the three free integration coefficients (after fixing the arbitrary overall normalization factor). The remaining condition is then treated as a complex residual function of $\omega$, the zeros of which are located using a numerical root-finding procedure. As an independent check, we have also implemented a version of the compound matrix method \cite{NG1985209}, obtaining consistent results.

In what follows, we first examine the spectrum of perturbations in the heat-flux sector alone, as discussed in Secs.~\ref{sec:results_T0} and \ref{sec:results_Q}, and then consider the coupled heat-flux and bulk-viscous sectors in Sec.~\ref{sec:BQ}.
For all results presented in this section, we adopt the EoS parameters specified in the end of Sec.~\ref{sec:eos} and the transport prescriptions discussed in Sec.~\ref{sec:transport} --- see also Table \ref{tab:transport}. We further define the characteristic gravitational timescale
\begin{equation}
    t_0 \equiv (G \rho_\mathrm{sat})^{-1/2} \approx 0.236 \,\mathrm{ms},
\end{equation}
associated with the nuclear saturation density $\rho_\mathrm{sat}$, and will often express transport coefficients in units of $t_0$.

\begin{table}
\centering
\begin{tabular}{lccc}
\hline
Channel & Transport coeff. & Constant param. \\
\hline
Bulk ($\Pi$)  & $\zeta$, $\tau_\Pi$ & $t_\zeta,\ t_\Pi$ \\
Shear ($\pi^{\mu\nu}$) &  $\eta$, $\tau_\pi$  & $t_\eta,\ t_\pi$ \\
Heat ($Q^\mu$)  & $\lambda_Q$, $\tau_Q$   & $t_\lambda,\ t_Q$ \\
\hline
\end{tabular}
\caption{Local transport coefficients and constant parameters for each dissipative channel. Equations \eqref{eq:transport_eta_zeta} and \eqref{eq:lambdaQ_prescription} specify the adopted prescriptions for $\{\zeta, \eta, \lambda_Q\}$ in terms of the parameters $\{\tau_\zeta, \tau_\eta, \tau_\lambda\}$. Equation~\eqref{eq:tau_prescription} relates the $\tau_i$'s to the constant parameters listed in the table.}
\label{tab:transport}
\end{table}

\subsection{Heat-flux sector at $T/\mu \to 0$} \label{sec:results_T0}

We begin this section by analyzing the $T\to0$ limit of perturbations with no bulk- or shear-viscous corrections ($\eta=\zeta=0$). In this case, the spectrum decouples as $\{\omega_n^\textrm{PF}\} \cup \{\omega_n^{(Q)}\}$ where, for each $n$, the Israel-Stewart theory admits one pair of perfect-fluid modes and one pair of thermal modes, while the Navier-Stokes limit admits a single mode in the thermal sector. For the fluid modes, $n$ denotes the number of nodes in the radial eigenfunction $\xi(r)$, while for the thermal modes it denotes the number of nodes in $Q(r)$, excluding possible zeros at the boundaries of the domain.

\begin{figure}[tbh] 
    \includegraphics[width=\linewidth]{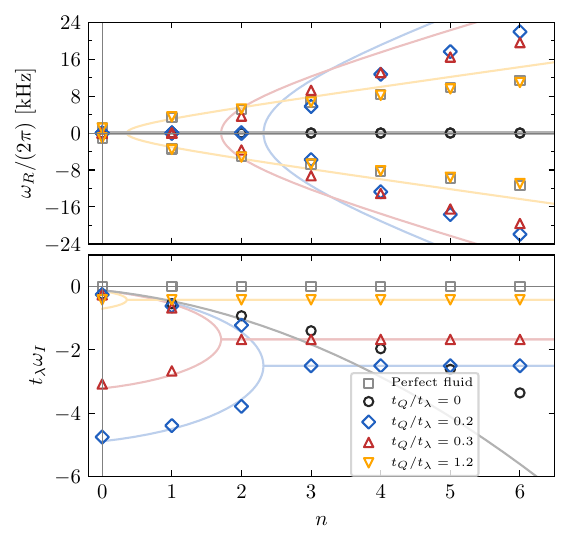}
    \caption{Real (upper panel) and imaginary (lower panel) parts of the mode frequencies as a function of the overtone number $n$ for a star with $C = 0.15$ and $T\approx0$ in the Israel-Stewart model, with $t_\lambda = 0.1 t_0$ and $t_Q/t_\lambda \in \{0, 0.2, 0.3, 1.2\}$. The fluid modes coincide with the perfect-fluid values at $T=0$ (shown as gray squares), while the thermal spectrum depends on the choice of transport coefficients. Colored curves correspond to the ansatz \eqref{eq:ansatzQ} for the thermal sector.}
    \label{fig:T0limit}
\end{figure}

Figure~\ref{fig:T0limit} shows the real and imaginary parts of the mode frequencies for a neutron star with compactness $C=0.15$, adopting $t_\lambda = 0.1 t_0$ and varying the ratio $t_Q/t_\lambda$, which is constrained to be $t_Q/t_\lambda \gtrsim 0.17$ by Eq.~\eqref{eq:causality_restricted} for this configuration. 
In the $T\to0$ limit, the perfect-fluid sector is unaffected by dissipation and therefore coincides for all configurations, whereas the thermal modes depend sensitively on the choice of transport coefficients.

In the Navier-Stokes ($t_Q = 0$) limit, the thermal modes have a purely imaginary frequency and become increasingly damped as $n$ increases. In the Israel-Stewart case, besides the Navier-Stokes-like thermal branch, there is an additional family of non-hydrodynamical modes, which, at $n=0$, have a frequency close to $-i t_Q^{-1}$ and remain purely damped for sufficiently small $n$. Above a critical value $n = n_c$, the two thermal branches merge into a pair of propagating modes whose real part grows with $n$, while the imaginary part is approximately constant.

These features of the thermal spectrum are broadly consistent with the flat-spacetime analysis of Sec.~\ref{sec:perturbations-homog}. In the rigid-fluid limit, the thermal sector is governed by the quadratic dispersion relation 
\begin{equation}
\tau_Q \omega^2 + i\omega + \lambda_Q (\partial_\mathrm{e}\alpha)_\mathrm{n} k^2 = 0,    
\end{equation}
which is precisely the dispersion relation associated with the Maxwell-Cattaneo equation \cite{cattaneo1958forme}, a causal generalization of the diffusion equation\footnote{The Maxwell-Cattaneo (or telegraph) equation for a diffusing quantity $n$ is $\tau \partial^2_t n + \partial_t n + D\nabla^2n=0$, with $D$ being the diffusion coefficient and $\tau>D$ a relaxation time that renders the theory hyperbolic and therefore causal.}. The two roots $\omega (k) = - i (2\tau_Q)^{-1} [1 \pm \sqrt{1-(k/k_c)^2}]$,
transition from purely damped to propagating at a critical wave number $k_c = [-4 \lambda_Q \tau_Q (\partial_\mathrm{e} \alpha)_\mathrm{n}]^{-1/2}$. For $k \gg k_c$, modes propagate with a (``second sound'') speed
\begin{equation} \label{eq:v2}
    v_2 = \pm \frac{1}{2\tau_Q k_c} = c_s \left| \frac{m_b(\partial_\mathrm{e}\alpha)_\mathrm{n}}{ (\partial_\mathrm{n}\alpha)_\mathrm{e}} \frac{t_\lambda}{t_Q} \right|^{1/2},
\end{equation}
where the second equality holds only for the transport prescriptions \eqref{eq:lambdaQ_prescription} and \eqref{eq:tau_prescription} adopted in this work. In this setup, $v_2$ is finite as $T\to 0$, since $\lim_{T\to 0}[(\partial_\mathrm{n}\alpha)_\mathrm{e}/(\partial_\mathrm{e}\alpha)_\mathrm{n}] = -\mu$.

Motivated by these considerations, one can construct a simple ansatz for the discrete thermal spectrum of the star by setting
\begin{equation} \label{eq:ansatzQ}
    \omega_n^\pm = -\frac{i}{2t_Q} \left[ 1 \pm \sqrt{1 - \left(\frac{k_n}{\bar{k}_c} \right)^2} \right], \quad k_n = \frac{\pi (n+1)}{R},
\end{equation}
where $\bar{k}_c$ is an effective value of the critical wavenumber, here taken to be $\bar{k}_c = k_c(R/2)$. The appearance of $t_Q$ instead of $\tau_Q$ in the denominator of Eq.~\eqref{eq:ansatzQ}---required to fit the data---can be interpreted as due to the implicit inclusion of a gravitational redshift factor to relate locally measured frequencies to those measured at infinity. 

The colored curves in Fig.~\ref{fig:T0limit} show the predictions of the ansatz \eqref{eq:ansatzQ}. Several features of the thermal spectrum are well reproduced by this simple expression, particularly the imaginary part of the propagating modes, which is consistent with the prediction $-(2 t_Q)^{-1}$ from \eqref{eq:ansatzQ}. The agreement is less accurate, however, for the real part of the propagating thermal modes, which is systematically overestimated.

\begin{figure}[tbh] 
    \includegraphics[width=0.98\linewidth]{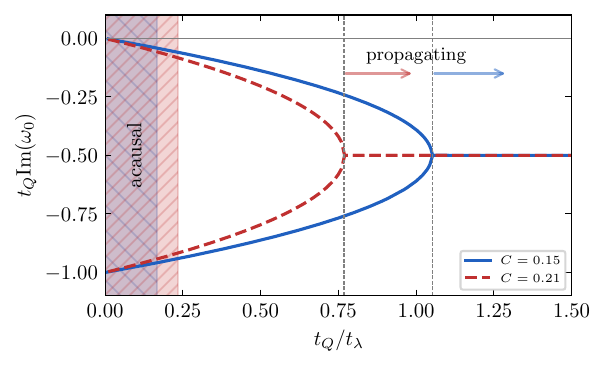}
    \caption{Imaginary part of the fundamental ($n=0$) thermal-mode frequencies, as a function of $t_Q/t_\lambda$, for stars with $C = 0.15$ and $C = 0.21$, in the $T\to 0$ limit, and in the Israel-Stewart model with $t_\lambda = 0.1 t_0$. The causality condition \eqref{eq:causality_condition} requires $t_Q/t_\lambda \gtrsim 0.17$ when $C=0.15$ and $t_Q/t_\lambda \gtrsim 0.23$ when $C = 0.21$. At a critical value of $t_Q$, the hydrodynamic and non-hydrodynamic thermal branches merge and transition into a pair of propagating modes.}
    \label{fig:n0T0}
\end{figure}

As $t_Q$ increases, both $k_c$ and, correspondingly, $n_c$ decrease, eventually reaching a regime in which even the fundamental ($n=0$) thermal modes become propagating, as illustrated in Fig.~\ref{fig:n0T0} and also by the $t_Q = 0.12\, t_\lambda$ case in Fig.~\ref{fig:T0limit}. This contrasts with the continuum limit, where, for any value of $\tau_Q$, there always exists a sufficiently small-$k$ regime in which the modes behave according to the long-wavelength approximation, Eq.~\eqref{eq:Taylor_omega}. The fact that the fundamental thermal modes can become propagating is tied to the fact that the stellar geometry imposes a finite wavelength to perturbations. This feature is captured by the ansatz \eqref{eq:ansatzQ}, which attributes to the $n=0$ mode a characteristic wavelength $2\pi/k_0 = 2R$. 

\subsection{Heat-flux sector at finite $T$} \label{sec:results_Q}

Let us now continue to focus on the heat-flux sector, i.e.~neglecting bulk and shear viscosity, and investigate how the conclusions obtained in the $T\to0$ limit are modified by the introduction of a finite temperature. We adopt the fiducial value $\chi = 0.1$ [cf.~Eq.~\eqref{eq:diffusive_equilibrium}], such that the temperature profile follows that of the chemical potential, albeit with an amplitude reduced by a factor of 10.
For a stellar compactness $C=0.15$, this choice corresponds to a central temperature of approximately $120$ MeV. Although this temperature is high---comparable to the largest temperatures reached in neutron-star mergers \cite{Perego:2019adq}---its effect on radial perturbations is modest.

The finite-$T$ corrections to the mode spectrum arise from two distinct sources. First, the finite temperature modifies the background stellar configuration and, consequently, the perturbation equations. Second, it breaks the block-triangular structure of the coefficient matrix ${\bf A}$ in Eq.~\eqref{eq:pert}, thereby coupling the fluid and thermal sectors and introducing dissipation into the fluid modes. Nevertheless, these effects remain small, scaling as $O(\chi^2)$. Indeed, all conclusions derived in Sec.~\ref{sec:results_T0} remain valid at $O(\chi)$.

\begin{figure}[tbh] 
    \includegraphics[width=0.98\linewidth]{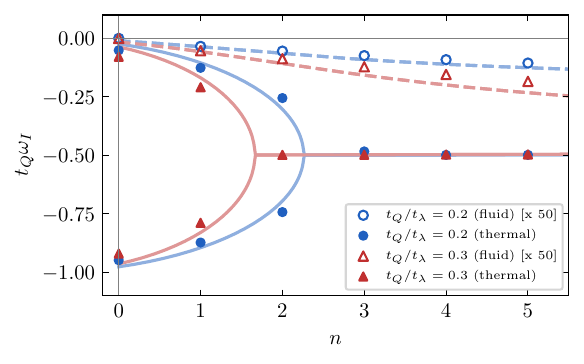}
    \caption{Imaginary part of the mode frequencies, rescaled by $t_Q$, as a function of the overtone number $n$, for a star with $C = 0.15$ and $\chi = 0.1$ in the Israel-Stewart model, with $t_\lambda = 0.1 t_0$ and $t_Q/t_\lambda \in \{0.2, 0.3\}$. Fluid (thermal) modes are shown as open (filled) symbols. For visualization purposes, the imaginary part of the fluid modes has been multiplied by a factor of 50. Colored curves correspond to the predictions of the ansatz \eqref{eq:ansatzT}. }
    \label{fig:Qonly}
\end{figure}

Figure \ref{fig:Qonly} shows the imaginary part of the fluid and thermal mode frequencies as functions of the overtone number $n$ for two values of the ratio $t_Q/t_\lambda$. The thermal spectrum is only mildly affected with respect to the previous section, while the fluid modes acquire a nonzero imaginary part. The damping of the fluid modes nevertheless remains much smaller than that of the thermal modes. One way to understand this separation of scales at small temperatures is to note from Eq.~\eqref{eq:Taylor_omega} that, in the small-$k$ regime,
\begin{equation}
    \frac{\Im[\omega_s^\pm]}{\Im[\omega_Q]} = \frac{\Gamma_s}{\Gamma_Q} \sim O(\chi^2).
\end{equation}
For $\chi = 0.1$, this scaling corresponds to a suppression of two orders of magnitude in the damping rate of the fluid modes relative to the thermal modes, consistent with the data shown in Fig.~\ref{fig:Qonly}.

The flat-spacetime analysis of Sec.~\ref{sec:perturbations-homog} also provides analytical guidance towards interpreting finite-$T$ results. Equation~\eqref{eq:determinant} is a sixth-order polynomial equation of the form
\begin{equation}
    \sum_{i=0}^6 c_i(k, {\bf y}) \omega^i(k) = 0,
\end{equation}
where the coefficients $c_i$ depend on the wavenumber $k$ and on the fluid variables, collectively denoted by ${\bf y}$, which are constant in the homogeneous-fluid analysis. This expression reduces to a fourth-order polynomial equation in the absence of bulk and shear viscosities, and its four roots define the dispersion relations for a pair of fluid (sound) modes and a pair of thermal modes. 

We find that the stellar spectrum is reasonably well approximated by: (i) replacing the fluid variables by representative background values, taken here as ${\bf y} \to {\bf \bar{y}} ={\bf y}(R/2)$, (ii) identifying the wavenumber with $k \to k_n = (n+1)\pi/R$, and (iii) introducing a redshift factor into the frequencies:
\begin{equation} \label{eq:ansatzT}
    \sum_{i=0}^6 c_i(k_n, {\bf \bar{y}}) \tilde{\omega}_n^i = 0, \qquad
    \omega_n \equiv e^{\bar{\Phi}/2}\tilde{\omega}_n.
\end{equation}
The full and dashed curves shown in Fig.~\ref{fig:Qonly} for the thermal and fluid modes, respectively, were obtained using this ansatz.
We note that this construction is not intended as a quantitative fit to the stellar eigenvalues, but rather as a qualitative interpretive tool that maps the discrete stellar spectrum onto the continuum dispersion relations. Its accuracy is limited by the replacement of the inhomogeneous stellar profile by midpoint values and by the use of a simple standing-wave quantization for $k_n$, both of which neglect the effects of the gravitational potential and the radial variation of thermodynamic quantities. Nevertheless, it captures the main qualitative features of the spectrum — the location of the damped-to-propagating transition, the asymptotic imaginary part of propagating modes, and the separation of scales between fluid and thermal damping rates — and provides a useful bridge between the flat-spacetime analysis of Sec.~\ref{sec:perturbations-homog} and the full numerical results.

\subsection{Bulk and heat-flux sectors} \label{sec:BQ}

As shown in Paper I \cite{Mendes:2025oib}, bulk viscous pressure can strongly affect the radial oscillation frequencies of neutron stars, especially for the fundamental $n=0$ mode, whereas for higher overtones its impact becomes comparable to that of shear viscosity. In this section, we complete our analysis of finite-temperature effects by considering the combined role of bulk viscosity and heat diffusion.

\begin{figure}[htb] 
    \includegraphics[width=0.98\linewidth]{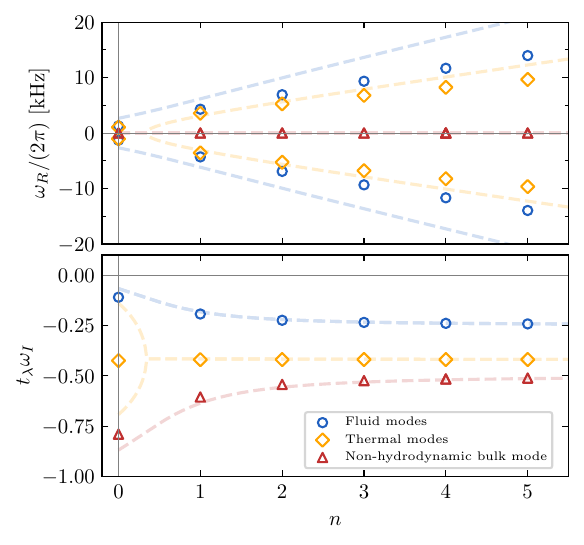}
    \caption{Real (upper panel) and imaginary (lower panel) parts of the mode frequencies as a function of the overtone number $n$ for a star with $C = 0.15$ and $\chi = 0.1$ in the Israel-Stewart model, with $t_\lambda = t_\zeta = t_\Pi = 0.1 t_0$ and $t_Q = 1.2 t_\lambda$. Colored curves correspond to predictions of the ansatz \eqref{eq:ansatzT}.}
    \label{fig:Bulk_and_Q}
\end{figure}

Figure \ref{fig:Bulk_and_Q} displays the radial mode spectrum for a star with $C = 0.15$ and $\chi = 0.1$, using transport coefficients with $t_\lambda = t_\zeta = t_\Pi = 0.1 t_0$ and $t_Q = 0.12 t_0$. Similar results are obtained for other, representative neutron-star compactnesses. The dashed curves correspond to the ansatz \eqref{eq:ansatzT}, motivated by the flat-spacetime analysis of Sec.~\ref{sec:perturbations-homog}.

The departure of the fluid modes from their perfect-fluid values is driven primarily by bulk viscosity, whereas heat diffusion mainly manifests through the appearance of two branches of thermal modes. For this choice of $t_Q/t_\lambda$, these thermal modes are propagating for all $n$. 
An analogous conclusion --- that the inclusion of heat flux gives rise to additional thermal modes, while the modes already present in its absence are weakly modified at low temperatures --- is verified in the coupled shear and heat-flux sectors.

\section{Conclusions} \label{sec:conclusions}

In this work we have extended the analysis of radial oscillations of dissipative relativistic stars within Israel-Stewart theory to include heat diffusion with a thermodynamically consistent finite-temperature equation of state. Together with the results of Paper I, where bulk and shear viscosity were considered, this completes the study of all dissipative degrees of freedom appearing in the Israel--Stewart framework. 
For the class of equations of state considered here, thermal effects remain parametrically suppressed in the low-temperature regime relevant for neutron stars. Consequently, heat transport leaves the ordinary fluid oscillations largely unchanged while generating a distinct thermal sector of excitations. In this limit, the latter is described by a causal diffusion process closely analogous to the Maxwell-Cattaneo equation, giving rise to both hydrodynamic and non-hydrodynamic thermal modes. 
The transition of these modes from diffusive to propagating behavior provides a finite-size realization of relativistic second sound, namely the propagation of thermal disturbances supported by the relaxation dynamics of the heat flux.

While the emergence of propagating thermal waves is interesting in its own right, the present analysis more importantly sheds light on the physical origin of the stellar oscillation spectrum. In Paper I it was observed that the spectrum of a dissipative star closely resembles the dispersion relations of perturbations around homogeneous equilibrium states. The inclusion of heat flow reinforces this correspondence. We find that the thermal branches appearing in the stellar spectrum reproduce, in discrete form, the characteristic structure of the Israel–Stewart dispersion relations, including hydrodynamic and non-hydrodynamic thermal branches and their transition into propagating thermal waves. 

At first sight, such a correspondence is far from obvious, as gravity and background inhomogeneities might be expected to substantially alter the dispersion relations originally derived for perturbations about an infinite, homogeneous medium. Instead, we have shown that a simple ansatz constructed directly from the homogeneous-fluid dispersion relations captures the main features of the stellar spectrum, including the approximate location of the diffusive-to-propagating transition and the characteristic damping scales of the thermal modes. Thus, beyond providing physical intuition, the homogeneous-fluid analysis may also serve as a useful semi-quantitative framework for understanding dissipative oscillations in compact stars. 

Interestingly, the finite stellar geometry and the resulting quantization of the spectrum introduce qualitatively new effects with respect to the continuum picture. In an infinite medium, sufficiently long wavelengths always probe the diffusive regime. In a finite star, however, the longest available wavelength is limited by the stellar radius, so the fundamental thermal mode need not lie in the hydrodynamic long-wavelength regime and may instead be propagating. As a result, depending on the transport coefficients, thermal-wave dynamics can become relevant already for the lowest stellar overtones.

Finally, we note that this work did not consider any effects due to radiative cooling through neutrino and photon emission. Such processes render the system open and play a central role in the long-term thermal evolution of compact stars. Nevertheless, we expect the qualitative mode classification identified here to remain robust, since it originates from the local structure of the hydrodynamic equations rather than from the specific mechanisms through which energy is exchanged with the environment. Extending the present framework to include radiative losses would constitute a natural next step towards more realistic descriptions of hot neutron stars and merger remnants.


\section*{Acknowledgments}

A.G.~acknowledges financial support from CAPES (Coordenação de Aperfeiçoamento de Pessoal de Nível Superior). G.S.D.~acknowledges CNPq (Conselho Nacional de Desenvolvimento Científico e Tecnológico), Grant No.~E-26/202.747/2018. R.M.~acknowledges financial support from CNPq as well as FAPERJ, Grant E-26/204.589/2024.  

\appendix



\bibliography{refs}
\bibliographystyle{apsrev4-2}

\end{document}